\documentclass[a4paper]{jpconf}
\usepackage{graphicx}
\usepackage{amsmath}

\begin{document}
\title{Adding CP to flavour symmetries}

\author{I. de Medeiros Varzielas}

\address{School of Physics and Astronomy, University of Southampton, Southampton, SO17 1BJ, U.K.}

\ead{ivo.de@soton.ac.uk}

\begin{abstract}
I propose the use of CP-odd invariants, which are independent of basis and valid for any choice of CP transformation,
as a powerful approach to study CP in the presence of flavour symmetries.
As examples of the approach I focus on Lagrangians invariant under $\Delta(27)$.
I comment on the consequences of adding a specific CP symmetry to a Lagrangian and distinguish cases where several $\Delta(27)$ singlets are present depending on how they couple to the triplets.
One of the examples included is a very simple toy model with explicit CP violation with calculable phases, which is referred to as explicit geometrical CP violation by comparison with previously known cases of (spontaneous) geometrical CP violation.
\end{abstract}

\section{Introduction}

This contribution to the proceedings of DISCRETE 2014 follows closely the layout of seminar I presented in the conference. I include here an expanded discussion of situations with $\Delta(27)$ singlets, including cases with explicit geometrical CP violation (first identified recently, in \cite{Branco:2015hea}). Some aspects discussed here are to appear also in a subsequent publication.

\subsection{The invariant approach}

I refer to the Invariant Approach (IA) to CP \cite{Bernabeu:1986fc} as an approach that starts by splitting the Lagrangian into $\mathcal{L}_{CP}$, a part that automatically conserves CP (e.g. kinetic terms, gauge interactions) and the remaining part $\mathcal{L}_{rem.}$:
\begin{equation}
\mathcal{L}=\mathcal{L}_{CP}+\mathcal{L}_{rem.} \,.
\end{equation}
The next steps are to
\begin{itemize}
\item Impose the most general CP transformations (that leave $\mathcal{L}_{CP}$ invariant).
\item Apply them and see if it restricts $\mathcal{L}_{rem.}$.
\end{itemize}
Only if the most general CP transformations restrict the shape of $\mathcal{L}_{rem.}$ can CP be violated.
An example of this type of restrictions is if the most general CP transformations force some coefficient to be real.

The IA is powerful because:
\begin{itemize}
\item Gets results just from the Lagrangian.
\item Independent of basis.
\item Shows relevant quantities for physical processes.
\end{itemize}

\subsection{The invariant approach for Standard Model leptons}

As a brief review of the IA, I apply it to a study of CP for Standard Model (SM) leptons. The mass Lagrangian is
\begin{equation}
\label{low}
 -\mathcal{L}_m=m_l  \overline{e}_L e_R +  \tfrac{1}{2} m_\nu \overline{\nu}_{L} \nu^{c}_L + h.c.\,,
\end{equation}
where $L= (e_L, \nu_L)$ stand for the left-handed neutrino and charged lepton fields in a weak basis; $e_R$ is the right-handed counterpart.

Due to the $SU(2)_L$ interactions (inside $\mathcal{L_{CP}}$), the most general CP transformations are:
\begin{eqnarray}
(CP) L (CP)^\dagger &=& i U \gamma^0 \mathcal{C} \bar{L}^T, \label{LCP1} \\
(CP) e_R (CP)^\dagger &=& i V \gamma^0 \mathcal{C} \bar{e}_R^T \,.
\label{LCP2}
\end{eqnarray}

I adopt a less precise notation that is more convenient to work with:
\begin{eqnarray}
L &\to& U L^* \,,\\
e_R &\to& V e_R^* \,.
\end{eqnarray}
I use this notation throughout, particularly as for simplicity I will mostly consider scalar fields in future sections, where the shorter notation is precise.

In order for $\mathcal{L}_m$ to be CP invariant, under eq(\ref{LCP1}), eq(\ref{LCP2}) the terms shown in eq(\ref{low}) go into the respective $h.c.$ and vice-versa:
\begin{equation}
U^\dagger  m_{\nu} U^* = m_{\nu}^*, \ \ \ \ 
U^\dagger m_{l} V = m_{l}^* \,.
\label{mlCP}
\end{equation}

Defining $H_\nu \equiv m_\nu m_\nu^\dagger$ and $H_l \equiv m_l m_l^\dagger$, I have:

\begin{equation}
U^\dagger  H_{\nu} U = H_{\nu}^*, \ \ \ \ 
U^\dagger H_{l} U = H_{l}^* \,.
\label{HlCP}
\end{equation}

At this stage I follow \cite{Bernabeu:1986fc} and build CP-odd invariants (CPI) by constructing combinations where $U$ and $V$ do not appear. First, I note that from eq(\ref{HlCP}), I can obtain
$\Tr ( H_\nu H_l ) = \Tr ( H_\nu H_l )^*$, which does not depend on $U$, $V$.
As the matrices are Hermitian, $\Tr ( H_\nu H_l )^* = \Tr ( H_\nu^T H_l^T ) = \Tr ( H_l H_\nu )^T = \Tr ( H_l H_\nu)$, concluding that for any CP transformations $U$, $V$, $\Tr (( H_\nu H_l ) - ( H_l H_\nu ))=0$ is required for CP conservation. Given that this is the trace of a commutator, this particular CPI automatically vanishes, meaning it is not very useful.
A more useful alternative is the necessary condition for CP conservation:
\begin{equation}
I_1 \equiv \Tr \left[H_\nu , H_l \right]^3 = 0\,,
\label{hhcube}
\end{equation}
valid for any number of fermion generations. For 3 generations (the SM case), it can be shown that eq(\ref{hhcube}) is a sufficient condition to have no Dirac-type CP violation in the lepton sector.

\section{The invariant approach and flavour symmetries}

One of the main points of my talk at the conference and of \cite{Branco:2015hea} is that the IA is very useful for analysing flavour symmetry models.
In order to illustrate this, I present some examples.

\subsection{Toy model with 4 couplings \label{sec:L4}}

\begin{figure}[h]
\begin{center}
\includegraphics[width=15 cm]{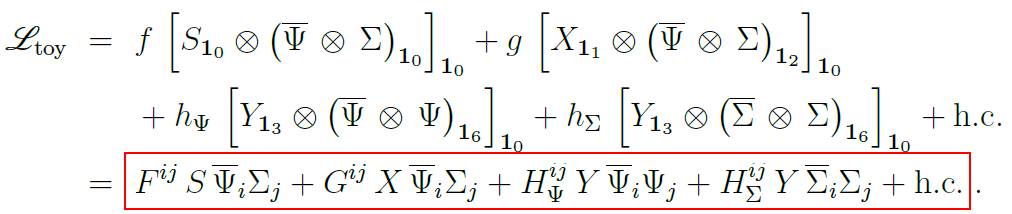}
\end{center}
\caption{Toy model Lagrangian from \cite{Chen:2014tpa}. \label{L_toy}}
\end{figure}

I start by considering a version of the toy model presented in section 3.1.1. of \cite{Chen:2014tpa}. As the aim here is to show the IA in action, I replace all fermions with scalars to avoid unnecessary complications. The Lagrangian (with fermions) as presented by \cite{Chen:2014tpa} is shown in figure \ref{L_toy}.

I refer to a similar Lagrangian (with scalars) as $\mathcal{L}_4$ (due to its 4 couplings):
\begin{equation}
\mathcal{L}_{4}=S \bar\Psi F \Sigma + X \bar\Psi G \Sigma + Y \bar\Psi H_\Psi \Psi + Y \bar\Sigma H_\Sigma \Sigma + h.c. \,,
\label{L4}
\end{equation}
where scalar fields $S$, $X$, $Y$ have just one generation, whereas scalar fields $\Psi$ and $\Sigma$ have $n$ generations, meaning $F$, $G$, $H_\Psi$, $H_\Sigma$ are $n \times n$ coupling matrices. In the original toy model, $\Psi$ and $\Sigma$ are fermions ( with $n=3$ generations) and I use the notation $\bar{\Psi}=\Psi^\dagger$, $\bar{\Sigma}=\Sigma^\dagger$ in eq(\ref{L4}) for easier comparison with the box in figure \ref{L_toy}.\footnote{I reconsider $\mathcal{L}_{4}$ in section \ref{L4again} making it invariant under a flavour symmetry, as done in \cite{Chen:2014tpa}.}

$\mathcal{L}_{4}$ is the $\mathcal{L}_{rem.}$ of this toy model and the most general $CP$ transformations (consistent with the respective $\mathcal{L_{CP}}$) are independent unitary transformations for each of the fields - phases for $S$, $X$, $Y$ and $n \times n$ unitary matrices $Q$ and $R$ for $\Psi$ and $\Sigma$:
\begin{eqnarray}
S &\to& e^{i s} S^* \,, \\
X &\to& e^{i x} X^* \,, \\
Y &\to& e^{i y} Y^* \,, \\
\Psi &\to& Q \Psi^*  \,, \\
\Sigma &\to& R \Sigma^* \,.
\end{eqnarray}

Imposing CP conservation requires $\mathcal{L}_{4}$ to be invariant under these, which implies that the terms displayed in eq(\ref{L4}) go into their $h.c.$ and vice-versa.
Starting with $Y \bar\Sigma H_\Sigma \Sigma$, I have
\begin{equation}
\mathcal{L}_{4} \supset  Y \bar\Sigma H_\Sigma \Sigma  + Y^* \bar\Sigma^* H_\Sigma^* \Sigma^* 
\end{equation}
and the relevant CP transformations
\begin{eqnarray}
Y &\to& e^{i y} Y^* \,, \\
\Sigma &\to& R \Sigma^*  \,,
\end{eqnarray}
act on $Y \bar\Sigma H_\Sigma \Sigma$:
\begin{equation}
Y \bar\Sigma H_\Sigma \Sigma \to e^{i y} Y^*  \bar\Sigma^* R^\dagger H_\Sigma R \Sigma^* \,.
\end{equation}
Comparing with the $h.c.$ I conclude that if $\mathcal{L}_{4}$ remains invariant under CP,
$e^{i y} R^\dagger H_\Sigma R = H_\Sigma^* $.
I repeat the procedure for the other 3 couplings and obtain the 4 relations
\begin{eqnarray}
e^{i s} Q^\dagger F R &=& F^* \,, \label{F}\\
e^{i x} Q^\dagger G R &=& G^* \,, \label{G}\\ 
e^{i y} Q^\dagger H_\Psi Q &=& H_\Psi^* \,, \label{HPsi} \\
e^{i y} R^\dagger H_\Sigma R &=& H_\Sigma^* \,. \label{HSigma}
\end{eqnarray}
These 4 relations are necessary and sufficient for $\mathcal{L}_{4}$ to conserve CP.
At this stage I build CPIs by combining the 4 relations to obtain equations where the general CP transformations cancel out, meaning they are independent of $s$, $x$, $y$, $Q$ and $R$. A relevant example is obtained by multiplying in order 1. the dagger of eq(\ref{F}), 2. eq(\ref{HPsi}), 3. eq(\ref{F}) and 4. the dagger of eq(\ref{HSigma}):
\begin{equation}
R^\dagger F^\dagger Q Q^\dagger H_\Psi Q Q^\dagger F R R^\dagger H_\Sigma^\dagger R \, e^{i (-s+y+s-y)} =  (F^\dagger H_\Psi F H_\Sigma^\dagger)^* \,,
\end{equation}
removing the phases and leaving unitary matrices only outside the product of couplings:
\begin{equation}
R^\dagger F^\dagger H_\Psi F H_\Sigma^\dagger R =  (F^\dagger H_\Psi F H_\Sigma^\dagger)^* \,.
\end{equation}
Doing the same with eq(\ref{G}) in steps 1. and 3. I would obtain similarly:
\begin{equation}
R^\dagger G^\dagger H_\Psi G H_\Sigma^\dagger R =  (G^\dagger H_\Psi G H_\Sigma^\dagger)^* \,.
\end{equation}
The remaining dependence on unitary matrix $R$ can be eliminated by taking the trace:
\begin{eqnarray}
\Tr \left[ F^\dagger H_\Psi F H_\Sigma^\dagger \right] = \Tr \left[ F^\dagger H_\Psi F H_\Sigma^\dagger \right]^*  &\to& \mathrm{Im}\Tr \left[ F^\dagger H_\Psi F H_\Sigma^\dagger \right] = 0 \,, \label{ImF} \\
\Tr \left[ G^\dagger H_\Psi G H_\Sigma^\dagger \right] = \Tr \left[ G^\dagger H_\Psi G H_\Sigma^\dagger \right]^* &\to& \mathrm{Im} \Tr \left[ G^\dagger H_\Psi G H_\Sigma^\dagger \right] = 0 \,. \label{ImG}
\end{eqnarray}

These conditions illustrate the power of the IA.
Simply from studying the CP properties of $\mathcal{L}_{4}$ I have found a set of necessary conditions for CP conservation (they are not necessarily sufficient).
The conditions are basis independent, valid for \emph{any} choice of CP transformation and for any number of generations $n$.
They also directly constrain quantities that are relevant for physical processes.

\begin{figure}[h]
\begin{center}
\includegraphics[width=15 cm]{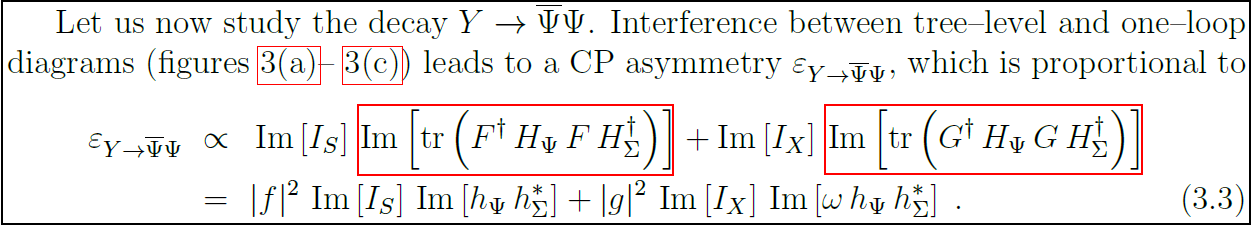}
\end{center}
\caption{Decay of $Y \to \bar\Psi \Psi$ as computed in \cite{Chen:2014tpa}. \label{Toy_decay}}
\end{figure}

In \cite{Chen:2014tpa}, the authors compute the CP asymmetry in the decay $Y \to \bar\Psi \Psi$, as shown in the text and equation displayed in figure \ref{Toy_decay}. By computing only a single CP asymmetry one might conclude that CP conservation can be obtained from a cancellation of the two quantities. Instead, by applying the IA to $\mathcal{L}_{4}$ I conclude that there are at least 2 independent necessary conditions for CP conservation, eq(\ref{ImF}), eq(\ref{ImG}).

\subsection{$\Delta(27)$ and adding CP} 

\subsubsection{$\Delta(27)$}

I discuss now the group theory of $\Delta(27)$ that is required for the remaining sections.
I define $\omega \equiv e^{i 2 \pi/3}$, the cyclic generator $c$ and diagonal generator $d$ of the group ($\omega^3=1$, $c^3=d^3=1$). There is an additional generator but it is not directly relevant for the discussion here. The group has irreducible representations that are either 1 or 3 dimensional - referred as singlets and triplets. The action of generators on singlets is simply multiplying them by a phase: $c 1_{ij}=\omega^i 1_{ij}$ and $d 1_{ij}=\omega^j 1_{ij}$, where $i, j = 0, 1, 2$ - there are 9 distinct singlets. In a convenient basis the action of the generators on a $3_{01}$ triplet $A=(a_1,a_2,a_3)_{01}$ or a $3_{02}$ triplet $\bar{B}=(\bar{b}_1,\bar{b}_2,\bar{b}_3)_{02}$ is:
\begin{equation}
c_{3_{0j}}=
\begin{pmatrix}
	0 & 1 & 0 \\
	0 & 0 & 1 \\
	1 & 0 & 0
\end{pmatrix}
 \,, \quad
c_{3_{01}}
\begin{pmatrix}
	a_1 \\
	a_2 \\
	a_3
\end{pmatrix}
=
\begin{pmatrix}
	a_2 \\
	a_3 \\
	a_1
\end{pmatrix} \,,
\end{equation}

\begin{equation}
d_{3_{01}}=
\begin{pmatrix}
	1 & 0 & 0 \\
	0 & \omega & 0 \\
	0 & 0 & \omega^2
\end{pmatrix} \,, \quad
d_{3_{02}}=
\begin{pmatrix}
	1 & 0 & 0 \\
	0 & \omega^2 & 0 \\
	0 & 0 & \omega
\end{pmatrix} \,.
\end{equation}
My nomenclature follows from the action of generators on triplets. The generator $d$ distinguishes the triplets $3_{01}$ and $3_{02}$ according to their subscripts, which are the powers of $\omega$ on the first two diagonal entries of the respective matrix. Hereafter I often refer to $3_{01}$ as the triplet representation and to $3_{02}$ as the anti-triplet representation. The cyclic generator acts equally on triplet and anti-triplet by cyclic permutation of the components.

The product of singlet with singlet leads to another singlet transforming as the sum of indices (modulo 3): $1_{ij} \times 1_{kl}$ transforms as $1_{(i+k) (j+l)}$. The product of triplet and anti-triplet gives a sum of all nine singlets. In the following it will be necessary to know how the singlets $1_{i0}$ and $1_{0j}$ are built from the product of triplet and anti-triplet.
$1_{00}$ is the trivial singlet transforming trivially under all generators and is formed from the $SU(3)$ contraction:
\begin{equation}
( A \bar{B} )_{00} \equiv (a_1 \bar{b}_1 + a_2 \bar{b}_2 + a_3 \bar{b}_3)_{00} \,.
\end{equation}

The $1_{i0}$ singlets are built as
\begin{eqnarray}
( A \bar{B} )_{10} &\equiv& (a_1 \bar{b}_1 + \omega^2 a_2 \bar{b}_2 + \omega a_3 \bar{b}_3)_{10} \,, \\
( A \bar{B} )_{20} &\equiv& (a_1 \bar{b}_1 + \omega a_2 \bar{b}_2 + \omega^2 a_3 \bar{b}_3)_{20} \,,
\end{eqnarray}
as acting with $c$ on $A$ and $\bar{B}$ leads to multiplication by $\omega$, $\omega^2$ respectively:
\begin{eqnarray}
( A \bar{B} )_{10} &\to& (a_2 \bar{b}_2 + \omega^2 a_3 \bar{b}_3 + \omega a_1 \bar{b}_1)_{10} \,, \\
( A \bar{B} )_{20} &\to & (a_2 \bar{b}_2 + \omega a_3 \bar{b}_3 + \omega^2 a_1 \bar{b}_1)_{20} \,.
\end{eqnarray}

In turn, the $1_{0j}$ are built as
\begin{eqnarray}
( A \bar{B} )_{01} &\equiv& (a_2 \bar{b}_1 + a_3 \bar{b}_2 + a_1 \bar{b}_3)_{01} \,, \\
( A \bar{B} )_{02} &\equiv& (a_1 \bar{b}_2 + a_2 \bar{b}_3 + a_3 \bar{b}_1)_{02} \,,
\end{eqnarray}
as acting with $d$ on $A$ and $\bar{B}$ leads to multiplication by $\omega$, $\omega^2$ respectively:
\begin{eqnarray}
( A \bar{B} )_{01} &\to& (\omega a_2 \bar{b}_1 + \omega^2 a_3 \omega^2 \bar{b}_2 +  a_1 \omega \bar{b}_3)_{01} \,, \\
( A \bar{B} )_{02} &\to& (a_1 \omega^2 \bar{b}_2 + \omega a_2 \omega \bar{b}_3 + \omega^2 a_3 \bar{b}_1)_{02} \,.
\end{eqnarray}

\subsubsection{Adding CP}
I consider now a specific $\Delta(27)$ invariant Lagrangian and study its CP properties.
The field content is triplet $A$, anti-triplet $\bar{B}$, and singlets $C$, $D$ (transforming respectively as $3_{01}$, $3_{02}$, $1_{10}$, $1_{01}$). The $\Delta(27)$ invariant Lagrangian for this field content contains one 3-field invariant between triplet, anti-triplet and each singlet:
\begin{equation}
\mathcal{L}_{CD}=y_c (A \bar B)_{20} C_{10} + y_d (A \bar B)_{02} D_{01} + h.c. \,.
\end{equation}
An additional $Z_N$ or $U(1)$ symmetry can be added to guarantee the absence of additional terms coupling $C$, $D$ to $A A^*$ or $\bar{B}^* \bar{B}$.
Focusing on the CP properties of $\mathcal{L}_{CD}$, I start by adding a specific CP transformation. A simple option is the trivial CP transformation $CP_1$, defined by the action on $A$, $\bar{B}$, $C$ and $D$:
\begin{eqnarray}
CP_1 A &=& A^*  = (a_1^*,a_2^*,a_3^*)_{02} \,, \\
CP_1 \bar{B} &=& \bar{B}^*= (\bar{b}_1^*,\bar{b}_2^*,\bar{b}_3^*)_{01} \,, \\
CP_1 C_{10} &=& C_{20}^* \,,\\
CP_1 D_{01} &=& D_{02}^* \,,
\end{eqnarray}
where $A^*$, $\bar{B}^*$, $C^*$, $D^*$ transform respectively as $3_{02}$, $3_{01}$, $1_{20}$, $1_{02}$ (reflected by the subscripts).

If I impose invariance under $CP_1$ on $\mathcal{L}_{CD}$,
the $y_c$ term which transforms to:
\begin{equation}
\to y_c (a_1^* \bar{b}_1^* + \omega a_2^* \bar{b}_2^* + \omega^2 a_3^* \bar{b}_3^*)_{20} C_{20}^* \,,
\label{CP1c}
\end{equation}
should become the $h.c.$, which features $y_c^*$:
\begin{equation}
y_c^* (a_1^* \bar{b}_1^* + \omega^2 a_2^* \bar{b}_2^* + \omega a_3^* \bar{b}_3^*)_{10} C_{20}^* \,.
\end{equation}
In addition to the conjugated coefficient, the expressions inside the parentheses are different, as denounced by their subscripts.
In turn, under $CP_1$ the $y_d$ term transforms into:
\begin{equation}
\to y_d (a_1^* \bar{b}_2^* + a_2^* \bar{b}_3^* + a_3^* \bar{b}_1^*)_{01} D_{02}^* \,,
\end{equation}
and comparing to its $h.c.$ with $y_d^*$
\begin{equation}
y_d^* (a_1^* \bar{b}_2^* + a_2^* \bar{b}_3^* + a_3^* \bar{b}_1^*)_{01} D_{02}^* \,,
\end{equation}
shows that apart from swapping $y_d$ to $y_d^*$ the expressions are the same.

A closer look at eq(\ref{CP1c}) shows that the transformed quantity is no longer invariant under $\Delta(27)$ (the subscripts do not add up to make a trivial singlet). One might state that, for this field content, $\Delta(27)$ is inconsistent with $CP_1$.
A more precise statement is that for $\mathcal{L}_{CD}$ to be invariant under both $\Delta(27)$ and $CP_1$ requires $y_c=0$ (and $y_d$ to be real) or alternatively, that insisting that $y_c \neq 0$ explicitly violates either $\Delta(27)$ or $CP_1$.
That $y_c$ is forced to vanish by adding a specific CP symmetry may appear drastic, but this is rather an usual consequence of adding symmetries to a Lagrangian. For example, one could also force $y_c=0$ in $\mathcal{L}_{CD}$ simply by having only the field $C$ transform non-trivially under an additional $Z_2$ symmetry.

One important point is that although imposing a specific CP transformation can force coefficients to vanish this needs not mean that CP violation occurs if those coefficients do not vanish. Indeed, $\mathcal{L}_{CD}$ with arbitrary $y_c$ and $y_d$ is CP conserving. I prefer to see this using the IA, and rewrite:
\begin{equation}
\mathcal{L}_{CD} = A_i Y_{10}^{ij} \bar{B}_j C + A_i Y_{01}^{ij} \bar{B}_j D + h.c. \,,
\end{equation}
with
\begin{equation}
Y_{10}=y_{c}
\begin{pmatrix}
1 & 0 & 0\\
0 & \omega & 0\\
0 & 0 & \omega^2
\end{pmatrix}
; \quad
Y_{01}=y_{d}
\begin{pmatrix}
0 & 1 & 0\\
0 & 0 & 1\\
1 & 0 & 0
\end{pmatrix} \,.
\label{Y01_Y10}
\end{equation}
Then I take the most general transformations
\begin{equation}
A \to U^* A^* ; \quad
\bar{B} \to V \bar{B}^* ;\quad
C \to e^{i p_{10}} C^*; \quad
D \to e^{i p_{01}} D^* \,,
\label{CP2s}
\end{equation}
and obtain the conditions for CP conservation
\begin{eqnarray}
U^\dagger Y_{01} V e^{i p_{01}} &=& Y_{01}^* \,, \label{Y2s1}\\
U^\dagger Y_{10} V e^{i p_{10}} &=& Y_{10}^* \,.
\label{Y2s2}
\end{eqnarray}
By building CPIs I conclude they are of the form
\begin{eqnarray}
\mathrm{Im}& \Tr& [ (Y_{01}^{\dagger} Y_{01})^{n_1} (Y_{10}^{\dagger} Y_{10})^{n_2} (Y_{01}^{\dagger} Y_{01})^{n_3} (...)] \,,
\label{IA2sG} \\
\mathrm{Im}& \Tr& [ (Y_{01} Y_{01}^{\dagger})^{n_1} (Y_{10} Y_{10}^{\dagger})^{n_2} (Y_{01} Y_{01}^{\dagger})^{n_3} (...)] \,,
\label{IA2sH}
\end{eqnarray}
where $n_i$ are positive integers.
These CPIs automatically vanish due to $\Delta(27)$, as $(Y_{01}^{\dagger} Y_{01})$, $(Y_{10}^{\dagger} Y_{10})$, $(Y_{01} Y_{01}^{\dagger})$ and $(Y_{10} Y_{10}^{\dagger})$ are proportional to the identity matrix\footnote{For the same reason, CPIs like $\mathrm{Im} \Tr [ (Y_{01}^{\dagger} Y_{01})^{n_1}  Y_{10}^\dagger Y_{01} (Y_{10}^{\dagger} Y_{10})^{n_2} Y_{01}^\dagger Y_{10}]$ give the same result as those in eq(\ref{IA2sG}).} with either $|y_{c}|^2$ or $|y_{d}|^2$.
The conclusion is that CP is conserved for any $y_c$, $y_d$.
Therefore, there must be at least one CP symmetry that leaves $\mathcal{L}_{CD}$ invariant regardless of arbitrary couplings (even though $CP_1$ does not).
One explicit example is:
\begin{equation}
U=
\begin{pmatrix}
1 & 0 & 0\\
0 & \omega^2 & 0\\
0 & 0 & 1
\end{pmatrix}
; \quad
V=
\begin{pmatrix}
1 & 0 & 0\\
0 & 1 & 0\\
0 & 0 & \omega^2
\end{pmatrix} ; \quad p_{10}= -2 \mathrm{Arg}(y_c) ; \quad p_{01}=-2 \mathrm{Arg}(y_d) \,.
\label{UVCD}
\end{equation}
This example is within the possibilities listed in \cite{Nishi:2013jqa} for CP transformations consistent with $\Delta(27)$ triplets.
While eq(\ref{UVCD}) applies to singlets $1_{01}$ and $1_{10}$, the reasoning based on the IA can be easily applied to any choice of two $\Delta(27)$ singlets.

\subsection{Additional singlets}

In the context of $\Delta(27)$ models with \emph{spontaneous geometrical CP violation}, meaning CP that is spontaneously broken with \emph{calculable phases} \cite{Branco:1983tn}, adding $\Delta(27)$ singlets coupling to triplet and anti-triplet was originally explored in \cite{deMedeirosVarzielas:2011zw}.
The goal was to obtain additional Yukawa couplings for SM fermions and the most promising choices considered the SM quark doublets as different singlets of $\Delta(27)$.\footnote{In \cite{deMedeirosVarzielas:2011zw} it was pointed out that one triplet of $\Delta(54)$ has the same scalar potential as one triplet of $\Delta(27)$, and options with irreducible representations of $\Delta(54)$ were also considered therein.}

Geometrical CP violation is a very interesting topic that continued to be explored after \cite{deMedeirosVarzielas:2011zw}.
Non-renormalisable terms in the scalar potential were considered in \cite{Varzielas:2012nn}, with focus on their effects on the calculable phases. 
The first viable model with SM quark doublets transforming as non-trivial $\Delta(27)$ singlets was realised in \cite{Bhattacharyya:2012pi}, featuring geometrical CP violation.
Subsequently it was shown in \cite{Varzielas:2013sla} how an additional symmetry can prevent the additional singlets from endangering the calculable phases and simultaneously explain the quark mass hierarchies. In \cite{Varzielas:2013eta} an extension of this type of model with the complete SM fermion sector was realised (see also \cite{Ma:2013xqa} for a different proposal for the leptons).

Geometrical CP violation was further explored in the context of multi-Higgs models with symmetries other than $\Delta(27)$ in \cite{Varzielas:2012pd, Ivanov:2013nla} (see also \cite{Varzielas:2013zbp}). The $\Delta(27)$ scalar potential for one triplet (invariant under $\Delta(54)$ \cite{deMedeirosVarzielas:2011zw}) was also analysed extensively with different approaches \cite{Holthausen:2012dk, Ivanov:2014doa, Fallbacher:2015rea}.

In \cite{deMedeirosVarzielas:2011zw, Varzielas:2012nn, Bhattacharyya:2012pi, Varzielas:2013sla, Varzielas:2013eta} CP is broken spontaneously, therefore the Lagrangian should conserve CP. As pointed out in \cite{Bhattacharyya:2012pi}, one must take care in adding extra singlets that couple to the triplet and anti-triplet as that may be incompatible with CP conservation.

One way to approach the constraints arising from adding singlets is by studying the outer automorphisms of the group, as discussed in \cite{Holthausen:2012dk} and also \cite{Chen:2014tpa}.
Alternatively, the reason why coupling more singlets to triplets may lead to CP violation becomes very clear in the IA: additional couplings enable more CPIs to be built. Eventually, adding an extra coupled singlet leads to a CPI that does not automatically vanish due to $\Delta(27)$, meaning one of 3 possibilities: $\Delta(27)$ is explicitly violated; CP is explicitly violated; or specific relations on the couplings are imposed. The last possibility is relevant in the context of \cite{deMedeirosVarzielas:2011zw, Varzielas:2012nn, Bhattacharyya:2012pi, Varzielas:2013sla, Varzielas:2013eta}, where one wants the symmetries to be broken spontaneously.
In analogy to $CP_1$ forcing $y_c=0$, one option that allows preserving both symmetries would be to have the couplings involving the triplets and some of the singlets vanish, and this is also understood clearly through the IA: the additional CPIs that do not vanish automatically can vanish due to the couplings.

Indeed, if one considers a set of singlets including two or three $1_{0j}$ singlets as in \cite{Bhattacharyya:2012pi}, the possible CP transformations on the triplets are already so constrained that one can not couple the triplets to even a single additional singlet $1_{ij}$ with $i \neq 0$ and preserve CP. Partly, this is why only $1_{0j}$ singlets were considered therein.

Strictly, the statement in \cite{Bhattacharyya:2012pi} that singlets $1_{ij}$ with $i \neq 0$ generate no coupling due to CP conservation is not mathematically rigorous. This statement is valid e.g. when $CP_1$ is imposed but not in general, as was pointed out recently in \cite{Fallbacher:2015rea}.
Nonetheless, for 3-field couplings between $\Delta(27)$ triplets and singlets (as in $\mathcal{L}_{CD}$), the physics of CP conserving situations - with at most 3 independent couplings - is contained in the choices considered in \cite{Bhattacharyya:2012pi}, which effectively corresponds to working in a basis where CP conservation is reflected on the Clebsch-Gordan (CG) coefficients being real.
An analysis of this issue also clarifies why the presence of any 2 singlets coupling in the manner of $\mathcal{L}_{CD}$ leads automatically to CP conservation.

\subsubsection{Changing basis}

Starting with just 1 singlet $S_{ij}$ and term $y_{ij} (A \bar{B})_{(-i)(-j)} S_{ij}$, I can always change the basis of $A$ such that the $(A \bar{B})_{(-i)(-j)}$ contraction looks like the corresponding contraction $(A \bar{B})_{(0)(-j)}$ in the original basis, which has real CG coefficients.
An explicit example is $y_{c} (A \bar{B})_{20} C_{10}$,
where the change of basis $(a_1, a_2, a_3) \to (a_1, \omega^2 a_2, \omega a_3)$ does precisely this:
\begin{equation}
(a_1 \bar{b}_1 + \omega a_2 \bar{b}_2 + \omega^2 a_3 \bar{b}_3)_{20} \to (a_1 \bar{b}_1 + a_2 \bar{b}_2 + a_3 \bar{b}_3)  \,.
\label{20basis}
\end{equation}
As far as the $y_{ij} (A \bar{B})_{(-i)(-j)} S_{ij}$ coupling is concerned it is equivalent to take a singlet in the set $1_{0j}$. 
Note that other couplings can distinguish the singlets, e.g. if $j=0$, terms $1_{00}$ and $1_{00}^2$ are $\Delta(27)$ invariants whereas the same does not apply for $1_{0j}$. But restricting ourselves to Lagrangian terms of the form of those in $\mathcal{L}_{CD}$ implies there are other symmetries that forbid such terms (such as the SM gauge group, in \cite{Bhattacharyya:2012pi}).

With 2 singlets, changing only the basis for $A$ may simply move the complex CG coefficients from one contraction into the other. $\mathcal{L}_{CD}$ is an explicit example of this as $(a_1, a_2, a_3) \to (a_1, \omega^2 a_2, \omega a_3)$ takes
\begin{eqnarray}
(a_1 \bar{b}_1 + \omega a_2 \bar{b}_2 + \omega^2 a_3 \bar{b}_3)_{20} &\to& (a_1 \bar{b}_1 + a_2 \bar{b}_2 + a_3 \bar{b}_3)  \,, \\
(a_1 \bar{b}_2 + a_2 \bar{b}_3 + a_3 \bar{b}_1)_{02} &\to& (a_1 \bar{b}_2 + \omega^2 a_2 \bar{b}_3 + \omega a_3 \bar{b}_1) \,.
\end{eqnarray}
But if one uses the change of basis:
\begin{eqnarray}
(a_1, a_2, a_3) &\to& (a_1, \omega^2 a_2, a_3) \,, \label{dbasis1} \\
(\bar{b}_1, \bar{b}_2, \bar{b}_3) &\to& (\bar{b}_1, \bar{b}_2, \omega \bar{b}_3) \label{dbasis2} \,,
\end{eqnarray}
then both singlets couple to triplets in $\mathcal{L}_{CD}$ with real CG coefficients:
\begin{eqnarray}
(a_1 \bar{b}_1 + \omega a_2 \bar{b}_2 + \omega^2 a_3 \bar{b}_3)_{20} &\to& (a_1 \bar{b}_1 + a_2 \bar{b}_2 + a_3 \bar{b}_3)  \,, \\
(a_1 \bar{b}_2 + a_2 \bar{b}_3 + a_3 \bar{b}_1)_{02} &\to& (a_1 \bar{b}_2 + a_2 \bar{b}_3 + a_3 \bar{b}_1) \,.
\end{eqnarray}
This change of basis takes $U$ and $V$ in eq(\ref{UVCD}) to the identity matrices of $CP_1$.

In a situation with 3 singlets coupling in the manner of $\mathcal{L}_{CD}$, the possibility of explicit CP violation depends on whether the freedom to change the basis of $A$ and $\bar{B}$ is enough to eliminate complex CG coefficients or not. 
Most choices of singlets can explicitly violate CP, but for 12 sets (out of 84 combinations) this is not possible. For these 12 sets there is at least one non-trivial element of $\Delta(27)$ which does not distinguish the 3 singlets. The special 12 sets can be identified in the notation I use here by summing the two generator indices over the 3 singlets - if both sums add up to 0 (modulo 3), an appropriate change of basis makes the CG coefficients real.
I demonstrate with the set $1_{00}$, $1_{10}$ and $1_{20}$:
\begin{eqnarray}
( A \bar{B} )_{00} &\equiv& (a_1 \bar{b}_1 + a_2 \bar{b}_2 + a_3 \bar{b}_3)_{00}\,, \\
( A \bar{B} )_{10} &\equiv& (a_1 \bar{b}_1 + \omega^2 a_2 \bar{b}_2 + \omega a_3 \bar{b}_3)_{10} \,, \\
( A \bar{B} )_{20} &\equiv& (a_1 \bar{b}_1 + \omega a_2 \bar{b}_2 + \omega^2 a_3 \bar{b}_3)_{20} \,,
\end{eqnarray}
The required basis change is not readily seen from the expressions, but noting the 3 singlets in the set are distinguished only by generator $c$, the basis change to eigenstates of $c_{3_{0j}}$:
\begin{eqnarray}
(a_1, a_2, a_3) &\to& (a_1+a_2+a_3, a_1+ \omega a_2+ \omega^2 a_3, a_1+\omega^2 a_2+ \omega a_3)/\sqrt{3}\,, \label{cbasis1} \\
(\bar{b}_1, \bar{b}_2, \bar{b}_3) &\to& (\bar{b}_1+\bar{b}_2+\bar{b}_3, \bar{b}_1+\omega^2 \bar{b}_2+\omega \bar{b}_3,\bar{b}_1+\omega \bar{b}_2+\omega^2 \bar{b}_3)/\sqrt{3} \,, \label{cbasis2}
\end{eqnarray}
takes the expressions to those of $(A \bar{B})_{0j}$, with real CG coefficients:
\begin{eqnarray}
(a_1 \bar{b}_1 + a_2 \bar{b}_2 + a_3 \bar{b}_3)_{00} &\to& (a_1 \bar{b}_1 + a_2 \bar{b}_2 + a_3 \bar{b}_3) \,, \\
(a_1 \bar{b}_1 + \omega^2 a_2 \bar{b}_2 + \omega a_3 \bar{b}_3)_{10} &\to& (a_2 \bar{b}_1 + a_3 \bar{b}_2 + a_1 \bar{b}_3) \,, \\
(a_1 \bar{b}_1 + \omega a_2 \bar{b}_2 + \omega^2 a_3 \bar{b}_3)_{20} &\to& (a_1 \bar{b}_2 + a_2 \bar{b}_3 + a_3 \bar{b}_1) \,.
\end{eqnarray}
The generalisation of the change of basis for sets of 3 singlets sharing a non-zero index is relatively straightforward for sets $1_{ij}$ sharing a fixed $i \neq 0$ (distinct only under generator $d$), where one has a diagonal change of basis (in analogy with eq(\ref{dbasis1}), eq(\ref{dbasis2})).
An explicit example is for singlets with a fixed $i=1$, with triplets contracting as:
\begin{eqnarray}
( A \bar{B} )_{20} &\equiv& (a_1 \bar{b}_1 + \omega a_2 \bar{b}_2 + \omega^2 a_3 \bar{b}_3)_{20} \,, \\
( A \bar{B} )_{21} &\equiv& (a_2 \bar{b}_1 + \omega a_3 \bar{b}_2 + \omega^2 a_1 \bar{b}_3)_{21} \,, \\
( A \bar{B} )_{22} &\equiv& (a_3 \bar{b}_1 + \omega a_1 \bar{b}_2 + \omega^2 a_2 \bar{b}_3)_{22} \,,
\end{eqnarray}
where a change to a basis with real CG is $(\bar{b}_1, \bar{b}_2, \bar{b}_3) \to (\bar{b}_1, \omega^2 \bar{b}_2, \omega \bar{b}_3)$.
For sets $1_{ij}$ sharing a fixed $j \neq 0$ (distinct only under generator $c$), the generalisation of the change of basis involves a mix of eq(\ref{cbasis1}), eq(\ref{cbasis2}) and the diagonal type similar to  eq(\ref{dbasis1}), eq(\ref{dbasis2}), or equivalently, reordering the eigenstates of $c_{3_{0j}}$ in eq(\ref{cbasis1}), eq(\ref{cbasis2}).
For sets not sharing an index, but for which the sum over indices both sum up to 0 (modulo 3) the change of basis is possible but requires an additional redefinition of one of the 3 singlets (in addition to the triplet representations).
Fortunately, the IA produces results that are basis independent, so for a given Lagrangian one can avoid checking whether basis changes that lead to real CG exist or not. 

Either using basis changes or the IA, the conclusion for Lagrangians with singlets coupling to triplet and anti-triplet in the manner of $\mathcal{L}_{CD}$ is the same.
There are 12 sets of 3 singlets that conserve CP, starting from $1_{00}$, $1_{01}$, $1_{02}$ and ending with $1_{20}$, $1_{21}$, $1_{22}$. The sets can be identified whenever the sum of both indices over the 3 singlets adds up to 0 (modulo 3), meaning that there is one non-trivial element of $\Delta(27)$ that does not distinguish the 3 singlets and it is then possible to choose that element to be the generator $c$ in another basis. As far as the 3-field couplings are concerned these 12 sets are equivalent through a change of basis to the choice with $i=0$, which is why this was the only set considered in \cite{Bhattacharyya:2012pi}.
For the other 72 choices of 3 singlets, or for 4 or more singlets, the complex CG coefficients can only be moved around by the change of basis, but not eliminated. In such situations, the coupling of the additional singlets to triplets is not allowed due to CP invariance of the Lagrangian, cf. \cite{Bhattacharyya:2012pi}.
 
A similar conclusion, based on an analysis of the automorphisms of $\Delta(27)$, was presented later in \cite{Chen:2014tpa}: that adding more than two non-trivial singlets (to a setting with just triplet representations) no longer allows a consistent CP transformation to be defined.\footnote{Strictly, how the singlets couple to triplet representations is very relevant, as discussed above cf. cases like the 8 sets of 3 non-trivial singlets that conserve CP automatically. Furthermore, this type of statement assumes all singlets have non-vanishing couplings to the triplets. This is not a spurious assumption as CP can be conserved even in settings with triplets and more than three non-trivial singlets, where triplet-decoupled singlets are still relevant due to coupling to other singlets. This can be a natural outcome if the vanishing couplings are enforced by a symmetry, which can be a specific CP symmetry as illustrated by $CP_1$ leading to $y_c=0$ for $\mathcal{L}_{CD}$.}

\subsubsection{$\Delta(27)$ and $\mathcal{L}_4$ \label{L4again}}

In \cite{Chen:2014tpa} a $\Delta(27)$ toy model with the trivial singlet and two non-trivial singlets was considered. This is actually the model illustrated here in figure \ref{L_toy}, which was the starting point for the scalar field Lagrangian $\mathcal{L}_4$ I used to exemplify the IA in section \ref{sec:L4}.
The authors of \cite{Chen:2014tpa} employed a $U(1)$ symmetry to restrict the allowed couplings and used the structures imposed by $\Delta(27)$ on the coupling matrices $F$, $G$, $H_\Psi$ and $H_\Sigma$ to compute the bottom row in figure \ref{Toy_decay}.

The field $S$ is associated with coupling matrix $F$ proportional to the identity so in their notation the $1_{0}$ singlet corresponds to the trivial singlet $1_{00}$ here. The fields $X$, $Y$ are associated to $G$ proportional to $Y_{01}$ and $H_\Psi$, $H_\Sigma$ proportional to $Y_{20}=Y_{10}^\dagger$ ($Y_{01}$ and $Y_{10}$ are shown in eq(\ref{Y01_Y10})), based on these couplings I identify that the singlets $1_{1}$ and $1_{3}$ in their notation correspond to singlets $1_{01}$ and $1_{20}$ here.
With these coupling matrices and the couplings $f$, $g$, $h_\Psi$ and $h_\Sigma$ defined in their Lagrangian as shown in figure \ref{L_toy}, calculating the CPIs in eq(\ref{ImF}), eq(\ref{ImG}) leads to:
\begin{eqnarray}
\mathrm{Im}\Tr \left[ F^\dagger H_\Psi F H_\Sigma^\dagger \right] &=&  \mathrm{Im} \left( |f|^2 h_\Psi h_\Sigma^* \right) \,, \label{ImFcalc} \\
\mathrm{Im} \Tr \left[ G^\dagger H_\Psi G H_\Sigma^\dagger \right] &=& \mathrm{Im} \left( \omega |g|^2 h_\Psi h_\Sigma^* \right) \,, \label{ImGcalc}
\end{eqnarray}
cf. figure \ref{Toy_decay}. Both CPIs depend on the phase $\mathrm{Arg}(h_{\Psi} h_{\Sigma}^*)$ (the relative phase between two arbitrary Lagrangian parameters). It is clear that no value of this phase can make both CPIs vanish so the conclusion is again that $\Delta(27)$ is explicitly violated, or CP is explicitly violated, or at least one coupling vanishes.
If I impose $f=0$ (or $g=0$) due to CP conservation, then $S$ (or $X$) decouples from the triplets $\Psi$, $\Sigma$ and CP can be conserved for specific values of the phase $\mathrm{Arg}(h_{\Psi} h_{\Sigma}^*)$. Note that the respective CP transformations for $f=0$ will differ from those for $g=0$. The triplet-decoupled singlet in either CP conserving case is still coupled to the other singlet scalars through quartic couplings that are unconstrained by the $U(1)$ symmetry,
$S S^\dagger X X^\dagger$, $S S^\dagger Y Y^\dagger$, $X X^\dagger Y Y^\dagger$.

\subsection{Explicit geometrical CP violation}

I propose now a toy model similar to $\mathcal{L}_{CD}$ but where the field content is reduced to contain only triplet $A$ and singlets $C_{10}$, $D_{01}$, and there are no $U(1)$ or $Z_N$ symmetries forbidding singlets from coupling to $(A A^*)$. Then the Lagrangian has terms:
\begin{equation}
\mathcal{L}_{A} = y_{c} (A A^*)_{20} C_{10} + y_{d} (A A^*)_{02} D_{01} + h.c. \,,
\end{equation}
In contrast with $\mathcal{L}_{CD}$ which has the same singlets, there is no $\bar{B}$.
The situation is to some extent similar to adding to $\mathcal{L}_{CD}$ the trivial singlet, in the sense that with couplings to a triplet in the manner of $\mathcal{L}_{A}$, the only pairs of singlets that automatically conserve CP are the 8 pairs including $1_{00}$, and the four pairs with 2 non-trivial singlets $1_{01}$, $1_{02}$; $1_{10}$, $1_{20}$; $1_{11}$, $1_{22}$; $1_{12}$, $1_{21}$ (where the sum over the 2 singlets of both indices adds up to 0 modulo 3).

Instead of dwelling further on basis changes
I use the IA to study the CP properties of $\mathcal{L}_{A}$. The most general transformations are the same from eq(\ref{CP2s}) (ignoring $\bar{B}$) and
the CP invariance conditions coming from $L_A$ are similar to eq(\ref{Y2s1}), eq(\ref{Y2s2}):
\begin{eqnarray}
U^\dagger Y_{01} U e^{i p_{01}}= Y_{01}^* \,, \\
U^\dagger Y_{10} U e^{i p_{10}}= Y_{10}^* \,.
\label{Y2sA}
\end{eqnarray}
Rather than trying to find unitary matrices that may not exist, it is often better to skip directly to building CPIs that do not depend on them. In this case a relevant CPI is:
\begin{equation}
I_{2} \equiv \mathrm{Im} \Tr (Y_{01} Y_{10}^\dagger Y_{01}^\dagger Y_{10}) \,,
\label{I2s}
\end{equation}
which has to vanish for CP conservation.
Using $Y_{01}$, $Y_{10}$ from eq(\ref{Y01_Y10}) I find:
\begin{equation}
I_{2} = \mathrm{Im} (3 \omega^2 |y_{c}|^2 |y_{d}|^2)
\label{I2sc}
\end{equation}
This means that CP can be \emph{explicitly} violated, in a minimal model with only 2 $\Delta(27)$ singlets.
But furthermore the IA also shows that in this model, CP is violated by 
a \emph{calculable phase} that is \emph{entirely determined by the symmetry of the Lagrangian} (and not by arbitrary parameters of the Lagrangian). The phases of the arbitrary $\mathcal{L}_{A}$ parameters $y_{c}$ and $y_{d}$ do not contribute, as shown very clearly in eq(\ref{I2sc}). This situation is directly comparable to the original definition of \emph{calculable phases} in \cite{Branco:1983tn}, where special cases with spontaneous CP violation were referred to as \emph{geometrical}. In analogy with the original definition, it is reasonable to refer to cases like this as \emph{explicit} geometrical CP violation.

Explicit geometrical CP violation was first identified in \cite{Branco:2015hea}. The model presented therein was not a scalar toy model but a physical multi-Higgs doublet model with scalars $h_{00}$, $h_{01}$ and $h_{10}$. It contains fermions $L$ (SM lepton doublets) and $\nu^c$ (SM singlet neutrinos) transforming under $\Delta(27)$ as triplet and anti-triplet. The neutrino Lagrangian is
\begin{equation}
\mathcal{L}_{3} = y_{00} (L \nu^c)_{00} h_{00} + y_{01} (L \nu^c)_{02} h_{01} + y_{10} (L \nu^c)_{20} h_{10} + h.c. \,.
\label{L3}
\end{equation}
CP is explicitly violated due to the presence of the 3 coupled singlets and the relevant CPI is naturally sensitive to all 3 couplings:
\begin{equation}
I_{3} \equiv \mathrm{Im} \Tr (Y_{00} Y_{01}^\dagger Y_{10} Y_{00}^\dagger Y_{01} Y_{10}^\dagger) \,,
\label{I3s}
\end{equation}
where $\Delta(27)$ imposes $Y_{00}$ proportional to the identity and I rename in eq(\ref{Y01_Y10}) $y_{c}=y_{10}$ and $y_{d}=y_{01}$ to match the notation from \cite{Branco:2015hea} used in eq(\ref{L3}).
 Then:
\begin{equation}
I_{3}=\mathrm{Im} (3 \omega^2 |y_{00}|^2  |y_{01}|^2 |y_{10}|^2) \,,
\end{equation}
showing CP is \emph{explicitly violated by a phase only originating from the group structure, and not from arbitrary couplings} - the arbitrary phases of $y_{00}$, $y_{01}$ and $y_{10}$ do not affect $I_3$.

\section{Conclusions}

The main conclusion to be drawn is that the invariant approach is a powerful method to study the CP properties of specific Lagrangians, particularly in the presence of flavour symmetries. The CP-odd invariants built from a Lagrangian do not require detailed knowledge of group theory and require relations between the couplings for the Lagrangian to conserve CP. One can then insert into the relevant CP-odd invariants the couplings that respect the flavour symmetry, and obtain a basis independent answer if CP is violated by those couplings.

I have illustrated the use of the invariant approach with several examples, mostly based on the $\Delta(27)$ symmetry. For a given Lagrangian I commented on the consequences of adding a specific CP symmetry.
I also clarified what are the possible outcomes when adding more $\Delta(27)$ singlets to different models, noting that it is relevant to distinguish how the singlets couple to triplets. For 
3 coupled singlets, a model with distinct triplet and anti-triplet (or two distinct triplets) can explicitly violate CP for 72 choices of 3 singlets; the other 12 choices 
lead to automatic CP conservation, which occurs when the 3 singlets  
are undistinguished by at least one non-trivial element of $\Delta(27)$ (this case includes 8 choices where the 3 singlets are non-trivial). In contrast, in a model with just one triplet the possibility for explicit CP violation exists already with 2 coupled non-trivial singlets. Finally, I used a simple toy model with a triplet and 2 coupled non-trivial singlets as an example of explicit geometrical CP violation, followed by the more realistic example with 3 singlets proposed in \cite{Branco:2015hea}.

\ack
This project is supported by the European Union's
Seventh Framework Programme for research, technological development and
demonstration under grant agreement no PIEF-GA-2012-327195 SIFT.
I thank the organisers of DISCRETE 2014 for hosting a very interesting conference, and also G. C. Branco, S. F. King for helpful discussions.

\end{document}